# Magnetization reversal using excitation of collective modes in nanodot matrices

Mehrdad Elyasi, Charanjit S. Bhatia, & Hyunsoo Yang

Department of Electrical and Computer Engineering, National University of Singapore, 4 Engineering Drive 3, Singapore 117576, Singapore

Correspondence and requests for materials should be addressed to H.Y. (eleyang@nus.edu.sg)

**The large arrays of magnetic dots are the building blocks of magnonic crystals and the emerging bit patterned media for future recording technology. In order to fully utilize the functionalities of high density magnetic nanodots, a method for the selective reversal of a single nanodot in a matrix of dots is desired. We have proposed a method for magnetization reversal of a single nanodot with microwave excitation in a matrix of magneto-statically interacting dots. The method is based on the excitation of collective modes and the spatial anomaly in the microwave power absorption. We perform numerical simulations to demonstrate the possibility of switching a single dot from any initial state of a 3 by 3 matrix of dots, and develop a theoretical model for the phenomena. We discuss the applicability of the proposed method for introducing defect modes in magnonic crystals as well as for future magnetic recording.**

The large arrays of magnetic dots are the basis of magnonic crystals and the emerging bit patterned media for future recording technology.[1-3] Resonance behaviors of such densely packed magnetic structure arrays have been a subject of recent intensified research due to their applications in microwave technology or signal transfer.[4-10] Due to their switchable ground state between a ferromagnet to antiferromagnet, magnetic dot lattices can be used as tunable magnonic crystals.[8,9] In addition, structural or magnetization non-uniformities in magnonic crystals lead to the appearance of defect modes, resulting in the modification of the spin-wave band-structure.[11,12] Magnonic crystals based on vortex arrays have been studied,[13-17] and it has been demonstrated experimentally that the state of a vortex matrix, where the diameter of each dot is 2 μm, is selectively formed under different frequencies of microwave excitation to achieve the lowest collective energy.[17]

Microwave assisted magnetization switching is based on the absorption of microwave excitation by the magnetic resonance of a magnetic structure. The absorbed energy reduces the dc magnetic field required for reversal of a high magnetic anisotropy material or structure. Microwave assisted switching in single and multi-domain magnetic structures has been demonstrated experimentally.[18-25] The theory behind microwave assisted switching has been assessed based on the instability of the resonance modes.[26-28] In addition, several numerical computations have been dedicated to study the effect of excitation and structural parameters in microwave assisted switching, as well as the applicability of this method for future magnetic recording technology.[27-32] Spin-transfer torque oscillators (STO)[33-37] are the potential microwave irradiators to be implemented in the write heads. The frequency and power of the microwave oscillation in STOs is tunable by varying their driving dc currents and the bias fields. For the implementation of microwave assisted switching, the excitation area should be decreased with the same rate by increasing the data storage density. Therefore, for achieving dots as small as several nano-meters, the write head should be able to irradiate microwave field focused down to the same area as the dot. Consequently, the STOs with the dimensions of several nm are required which is beyond the current fabrication capabilities.

In order to fully utilize the functionalities of high density magnetic nanodots for magnonic crystals and future recording applications, a method for the selective reversal of a single nanodot in a matrix of dots is desired. For this purpose, we propose a selective magnetization switching method with microwave excitation based on dispersion in the power absorption of the collective magnetic resonance modes. Numerical simulations performed by object oriented micromagnetic framework (OOMMF) are used to support the idea. In addition, a theoretical model of the system is developed to have a better insight on the physical interpretation of the phenomenon. We also assess the applicability of such method for applications.

## Results

**Numerical analysis.** Microwave excitation of a magnetic structure leads to magnetization dynamics which is a non-linear process. If absorption of the microwave power is more than Gilbert or spin wave damping, the magnetization cone angle grows from its initial stable condition while precessing. In a certain condition, the magnetization overcomes the energy barrier and falls into the other energy well. The magnetization dynamic process occurs by coupling the stored microwave power in the main precessional mode to the possible spin wave modes of the structure. A recent microwave assisted switching proposal is to excite a single dot with microwave in the presence of a dc field by considering the effect of the magnetizations of neighboring dots as sources of stray field.[38] A magnetic dot array, which is interacting magneto-statically, exerts a dispersion relation based on the lattice, array, and the dot dimensions.[5,8] Therefore, absorption of microwave excitation with different frequencies leads to different spin wave modes of the dot array structure. This feature provides an opportunity to investigate the microwave assisted switching process with exciting a matrix of magnetic dots (Fig. 1(a)). If a dc field is applied to a region of a dot array containing a N×N matrix of dots, there is a step change in the Zeeman energy of that region relative to the surrounding area. The N×N dot matrix has localized collective modes originated from the magneto-static interactions. Each of the $2^N$ possible oscillation modes in such matrix has a specific resonance frequency and absorption characteristic. An anomaly in the power absorption spectrum among the different oscillation modes causes nonuniformity in the oscillation amplitude among the dots. Therefore, it is expected that by exciting the dot matrix by microwave fields with different frequencies, different oscillation amplitude patterns emerge, consequently resulting in reversal of the magnetization of dots into different patterns.

In the following discussion, we present the results of the numerical simulations using OOMMF. Figure 1(b) demonstrates the reversal occurrence of different patterns with respect to the microwave excitation frequency and duration for a 3×3 matrix of 10 nm×10 nm×10 nm dots. The inter-distance ($w_d$) between the dots in the $x$ and $y$ direction is 10 nm. The magnetic dots have uniaxial crystalline anisotropy in the $z$ direction, and the anisotropy constant $K_u$ is assumed to be $10^6$ J/m$^3$.[39] The saturation magnetization $M_s$, the Gilbert damping constant $\alpha$, and the gyromagnetic ratio $\gamma$, are set to $10^6$ A/m, 0.005, and $2.2\times10^5$ m/(A·s), respectively. The dc field is $\vec{H}_{dc} = -H_{dc}\hat{z}$, where $H_{dc}$ is 14.5 kOe. The microwave field is $\vec{h}_{mw}(t) = h_{mw}sin(\omega t)\hat{x}$, with $h_{mw}$ equal to 400 Oe and $\omega = 2\pi f$. The initial magnetization direction for all the dots is assumed to be aligned to the z direction. The frequency $f$ of the microwave field has been swept from 250 MHz to 12.5 GHz with a step of 250 MHz. For each of the $f$ values, the magnetization switching pattern has been tracked up to 15 ns of excitation duration. Overall 26 different switching patterns are achieved with different excitation frequencies and durations with the above field values. All the emerging switching states are demonstrated as red color in Fig. 1(b). To visualize better, the frequencies and durations required for achieving only 10 out of 26 possible switching patterns have been depicted and color-coded in Fig. 1(c).

Different initial states of the magnetization in dots result in different mode dispersions. Therefore, different switching patterns and sequences are expected for different initial states depending on the excitation frequencies and durations. Figure 2(a) shows the switching behavior when two corners of the matrix are initially magnetized in the –z direction (marked in red). 18 different switching patterns are achieved for this initial state and four representative patterns have been depicted in Fig. 2(a). In order to have a better insight in the relation between the absorption of each dot in different frequencies and the corresponding switching state, squares of the magnetization ($M_z^2$) in the z direction of all the 9 dots have been plotted against the microwave excitation frequencies in Fig. 2(b) at 400 ps, which is well before switching of any dot. In Figs. 2(b) and 2(d), the red color (high $M_z^2$ value) indicates there is no oscillation, while the blue color (low $M_z^2$ value) represents high oscillation amplitudes. As shown in Fig. 2(b) the oscillation amplitude is maximum (corresponding to the lowest $M_z^2$ value) in the central dot (dot number 5). It is observed in Fig. 2(a) that for the frequencies indicated by black arrows in Fig. 2(b) the switching of only the central dot occurs (marked in light blue). As an example of the sequential emergence of different switching states, the reversal pattern sequence and the temporal evolution of $M_z^2$ at $f = 10.75$ GHz are depicted in Fig. 2(c) and (d), respectively. It can be seen in Fig. 2(b) that at $f = 10.75$ GHz, the dot number 5 has the maximum oscillation amplitude (low $M_z^2$ value), dots 3 and 7 have an intermediate oscillation amplitude, and the oscillation for dots 2 and 8 is the lowest. As shown in Fig. 2(c-d), the sequence of switching at this frequency occurs in this order (dot 5 → 3,7 → 2,8). It is worth noting that at this frequency, the dots 4 and 6 equilibrate in a non-growing resonance mode (Fig. 2(d)). Figure 2(e) shows the pattern of the oscillation power ($P_o = 1-M_z^2/M_s^2$) for a relatively lower microwave field, $h_{mw} = 50$ Oe. The characteristics of oscillation patterns with frequency are similar to that in Fig. 2(b). The similarity confirms that the reversal patterns are initiated from the excited collective modes. An overall shift to higher frequencies of the patterns is due to a lower oscillation angle, and thus a higher anisotropy field. Figure 2(f) shows the phases of the oscillations in the dots with respect to the central dot (dot 5), at a particular time ($t = 3$ ns). A spatial variation of the phase ($\varphi_o$) confirms the existence of multiple collective modes at different frequencies.

The amplitude of dc ($H_{dc}$) and microwave ($h_{mw}$) fields affects the switching behavior. The value of $H_{dc}$ determines the frequency spectrum of the possible collective modes of the oscillations, whereas $h_{mw}$ affects the change of the magnetization oscillation of each mode. The total effect is the summation of changes of all the possible modes. Figure 3(a) shows the total number of possible switching patterns ($N_s$) with various $H_{dc}$ from the initial state of all dots tilted $12°$ from the $z$ direction. For $H_{dc}$ greater than a threshold value, $N_s$ is close to 50. Figure 3(b) shows frequencies required for only reversing the central dot (dot 5) out of the matrix for different dc fields. A redshift with increasing $H_{dc}$ up to 16 kOe is observed. This decrease in frequency is in conjunction with the relation of a single mode frequency and $H_{dc}$. For values of $H_{dc} > 16$ kOe, the competition among modes results in new values of frequency required for the reversal of only the central dot. $N_s$ versus $h_{mw}$ in Fig. 3(c) shows a peak at $h_{mw}$ = 0.4 kOe, whereas the frequencies required for reversal of only the central dot do not show any strong feature in Fig. 3(d), as the effect of $h_{mw}$ on the collective reversal process is highly nonlinear.

Despite the ability to achieve different switching patterns from a particular initial state, it is not possible to change the magnetization from an arbitrary initial pattern to any final pattern in one step of switching. As the absorption is the maximum in the central dot, we focus on the ability of reversing only dot 5 from different initial states of the matrix. The total number of initial states for a 3×3 matrix is $2^9$, in which each dot can be aligned to the $-z$ or $+z$ direction. Since $\vec{h}_{mw}$ oscillates in the $x$ direction, there are two mirror symmetry axes of (100) and (010). In consequence, there exist only 168 independent initial states. Excluding the initial patterns with dot 5 being down ($-z$ direction), 84 states remain. Since, there is no difference in applying microwave field in the $x$ or $y$ direction, only 51 initial states should be considered. Figure 4(a) shows the microwave excitation frequencies and durations in which the reversal of only dot 5 occurs. It is observed that for *all* the independent initial states, the reversal of sole dot 5 is possible. Although we have concentrated on the reversal of only the center dot, achieving different reversal patterns from different initial states is possible. Figure 4(b) shows the ratio of the realized reversal patterns ($N_s$) out of the possible patterns ($N_p$) depending on the initial states. It can be seen that for limited number of initial states we can achieve $N_s/N_p$ equal to unity. As expected, the initial states with strong symmetry result in small values of $N_s/N_p$.

The ability of reversal of only the center dot is not limited to 3×3 matrices of dots. Figure 5 shows some of the possible reversal patterns for a 7×7 matrix of dots from the initial state of all the dots being up. Among the depicted reversal patterns, the frequencies in which the reversal of only the center dot happens are depicted in dark red color.

**Theoretical model.** We assume that the N×N dot matrix under excitation is embedded in a very large array of dots. Each dot has the planar dimension of $w \times w$ and the thickness of $h$. The distance between dots in the $x$ and $y$ direction is $w_d$. The shape of the dots is not critical, and we can expand the approach to cylindrical or any other shapes. We assume a uniaxial crystalline anisotropy to the z-direction with the anisotropy constant of $K_u$. The magnetization of the dots is defined as

$$\left[\vec{M}\right]_{N \times N} = \left[\vec{M}_0\right]_{N \times N} + \left[\vec{m}(t)\right]_{N \times N}, \tag{1}$$

where $\left[\vec{M}_0\right]_{N \times N}$ is the matrix of initial magnetization states, and $\left[\vec{m}(t)\right]_{N \times N}$ is the matrix of dynamic magnetization as a function of time, $t$. For the chosen material, the initial magnetization matrix is

$$\left[\vec{M}_0\right]_{N \times N} = M_s \left[S\right]_{N \times N} \hat{z}, \tag{2}$$

where $\left[S\right]_{N \times N}$ is 1 or -1. The oscillation of each mode can be described by their complex time varying amplitudes $\vec{C}_i(t)$ as

$$\left[\vec{K}_i\right]_{N \times N} = \vec{C}_i(t) \left[K_i\right]_{N \times N}, \tag{3}$$

where $\left[K_i\right]_{N \times N}$ is a binary permutation of the N×N matrix elements (dots), and $i$ is from 1 to $N_m$. Depending on the initial state, $N_m$ can be different from 1 to $2^N$. Therefore, the total magnetization dynamic matrix is $\left[\vec{m}(t)\right]_{N \times N} = \sum_{i=1}^{N_m} \left[\vec{C}_i(t)\left[K_i\right]_{N \times N} + c.c.\right]$, where c. c. denotes the complex conjugate.

For analyzing the dynamics of each mode, we use the Landau-Lifshitz-Gilbert (LLG) equation,

$$\frac{d\vec{C}_i(t)}{dt} = \gamma\left(\vec{H}_{eff,i} \times \vec{C}_i(t)\right) - \frac{\alpha\gamma}{M_s}\left(\left(\vec{H}_{eff,i} \times \vec{C}_i(t)\right) \times \vec{C}_i(t)\right), \quad (4)$$

where $\vec{H}_{eff,i}$ is the effective field for each mode and can be written as

$$\vec{H}_{eff,i} = \vec{H}_{dc,eff} + \vec{h}_{mw} + \vec{H}_{d,i} + \vec{H}_A. \quad (5)$$

$\vec{H}_{dc,eff} = \vec{H}_{dc} + \vec{H}_{OB}$, $\vec{H}_{OB}$ is the magneto-static field from the dots outside the $N \times N$ matrix, $\vec{H}_A = \frac{2K_u}{\mu_0 M_s^2}\left(\vec{C}_i(t) \cdot \hat{z}\right)\hat{z}$ is the crystalline anisotropy field, and $\vec{H}_{d,i}$ is the demagnetizing field for the mode $i$, which can be written as

$$\vec{H}_{d,i} = -\sum_{p,q}\sum_{m,n} \hat{N}\left(\vec{r}_{pq} - \vec{r}_{mn}\right) \cdot \vec{M}_{0,pq}, \quad (6)$$

where $\hat{N}\left(\vec{r}_{pq} - \vec{r}_{mn}\right)$ is the demagnetizing tensor, $\vec{r}_{pq}$ ($\vec{r}_{mn}$) is the center of the dot in the row $p$ (m) and the column $q$ (n). The first summation runs over $m$ and $n$ from 1 to $N$, while $p$ and $q$ run over only the elements that are included in the mode $i$; in other words for each $m,n \in 1,2,...,N$ such that $K_{i,mn} = 1$. To calculate the demagnetizing tensor, we can use[40]

$$\hat{N}(\vec{r}) = D(\vec{r}) \otimes \Lambda(\vec{r}), \quad (7)$$

where $D(\vec{r})$ is the shape function of the dot, and $\Lambda(\vec{r})$ is the demagnetizing tensor for a dipole $D(\vec{r}) = \delta(\vec{r})$ which can be written as

$$\Lambda(\vec{r}) = \frac{1}{4\pi r^5}\begin{bmatrix} r^2 - 3x^2 & -3xy & -3xz \\ -3xy & r^2 - 3y^2 & -3yz \\ -3xz & -3yz & r^2 - 3z^2 \end{bmatrix}. \quad (8)$$

The resonance frequency of each mode can be derived from equation (4) by neglecting the Gilbert damping and assuming $h_{mw} = 0$. By assuming $\vec{C}_i(t) = \vec{c}_i(t)e^{j\omega t}$, we can write

$$\omega_i = \gamma\left(\left(\left(N_{z,i} - N_{y,i}\right)M_s - \left|\vec{H}_A\right| + H_{dc}\right)\left(\left(N_{z,i} - N_{x,i}\right)M_s - \left|\vec{H}_A\right| + H_{dc}\right)\right)^{1/2}, \tag{9}$$

where $N_{x,i}$ ($N_{y,i}$ and $N_{z,i}$) is the first (second and third) element of $\sum_{p,q}\sum_{m,n} \hat{N}(\vec{r}_{pq} - \vec{r}_{mn}) \cdot [1\ 1\ 1]/\sqrt{3}$ for the mode $i$.

Exciting the matrix of dots with microwave results in an absorption of power by possible modes, in which the resonance frequency ($\omega_i$) is different for the various modes. Assuming $\vec{h}_{mw} = h_{mw} e^{-j\omega_{mw}t} \hat{x}$, equation (4) results in the steady susceptibility tensor $\hat{\chi}(\omega)$,

$$\hat{\chi}(\omega) = \gamma M_s \begin{bmatrix} j(\omega_i - \omega_{mw}) & \gamma\left((N_z - N_y)M_s - \left|\vec{H}_A\right| + H_{dc}\right) - \alpha j(\omega_i - \omega_{mw}) \\ \gamma\left((N_x - N_z)M_s + \left|\vec{H}_A\right| - H_{dc}\right) + \alpha j(\omega_i - \omega_{mw}) & j(\omega_i - \omega_{mw}) \end{bmatrix}^{-1}. \tag{10}$$

Therefore, we can write the stationary values of $\vec{c}_i$ as $\begin{bmatrix} c_{i,x} \\ c_{i,y} \end{bmatrix} = \hat{\chi}(\omega) \begin{bmatrix} h_{mw,y} \\ h_{mw,x} \end{bmatrix}$. Finally, we define the total stationary value of dynamic as $[\vec{c}_T]_{N\times N \times 2} = \sum_{i=1}^{N_m} \vec{c}_i [K_i]_{N \times N}$. Even though $[|\vec{c}_T|]_{N \times N}$ does not infer any information about the growth of the magnetization dynamics, it is a good measure which indicates the relative oscillation amplitude among the dots of the matrix, where the possible reversal is expected to be initiated from the dots with the largest absorption (Fig. 2).

As discussed before, the independent initial states of the magnetization ($[\vec{M}_0]_{N \times N}$) that should be considered are less than the total possible states of $2^N$, due to the symmetries induced by $\vec{H}_{dc}$ (infinite mirror planes in the $xy$ plane) and $\vec{h}_{mw}$ (mirror axes along (1 0 0) and (0 1 0)). In addition, if $\vec{h}_{mw}$ is linearly polarized along the $x$ or y axis, or circularly polarized, some states can be assumed as degenerate and only one of them is analyzed. As mentioned before, for the case of $N = 3$, there are 51 independent initial states where the central dot is magnetized to the z direction. Using the above analytical method, we calculate $\left[\left|\vec{c}_T\right|\right]_{3\times 3}$ for all the independent initial states ($\left[\vec{M}_0\right]_{3\times 3}$), and extract the frequencies in which the central dot has the highest oscillation amplitude ($\left|\vec{c}_T\right|_{2,2} \geq \left|\vec{c}_T\right|_{m,n} \forall m,n \in 1,2,3$). Figure 6(a) shows that there are several frequencies for most of the independent $\left[\vec{M}_0\right]_{3\times 3}$ with a combination of $\left|\vec{H}_{dc,eff}\right|$, in which the absorption of the central dot is maximum. If the central dot has the highest absorption amplitude for a particular frequency, it indicates the possibility of the sole reversal of the central dot with the growth of the collective modes. However, dynamic consideration of the collective modes can modify the reversal process as we discuss below. The microwave excitation is coupled to the collective oscillation modes and results in their growth and consequently reversal of the magnetization of the dots. For deriving the temporal behavior of each of the collective modes,[9] we use the circularly polarized basis for each of the modes where we can write $\vec{C}_i(t) = \vec{m}_i c_i^{'}(t)$ and $\vec{m}_i = (-i, 1, 0)$. Since the coordination basis is unique for each of the modes, scattering between the modes is prohibited. Due to the assumption of small oscillations, $H_A$ becomes constant and can be included in $H_{dc}$. After some algebraic simplification and keeping only the linear terms in $c_i^{'}(t)$, we can write

$$\frac{dc_i^{'}}{dt} = -i\Omega_i c_i^{'} - iR_i c_i^{'*} - \Gamma_i c_i^{'} + \xi_i(t) \tag{11}$$

where

$$\Omega_i = \gamma\mu_0 H_{dc} - \gamma\mu_0 H_{d,i}^z - \gamma\mu_0 \left(\frac{H_{d,i}^x + H_{d,i}^y}{2}\right) \tag{12}$$

$$R_i = -\gamma\mu_0(1+i\alpha)\left(\frac{H_{d,i}^y - H_{d,i}^x}{2}\right)$$

$$\Gamma_i = \alpha\Omega_i.$$

The term, $\xi_i(t)$ containing the coupling of the microwave excitation to the oscillation mode $i$ can be written as

$$\xi_i(t) = i\gamma\mu_0 M_s(\vec{m}_i \cdot \vec{h}_{mw}). \tag{13}$$

In the circularly polarized basis for the mode $i$, the linearly polarized $\vec{h}_{mw}$ along the $x$ direction is equal to $\left(h_{mw}e^{j(\omega_{mw}-\omega_i)t},0,0\right)$. By writing the same equation as equation (11) for $c_i^*$ and some algebraic simplifications, the second order non-homogenous differential equation for $c_i$ is

$$\frac{d^2 c_i'}{dt^2} - 2\Gamma_i \frac{dc_i'}{dt} + \lambda_i c_i' = f_i(t) \tag{14}$$

where $\lambda_i = \Omega_i^2 + |R_i|^2 + \Gamma_i^2$, and $f_i(t) = -i\Omega_i\xi_i + \Gamma_i\xi_i - iR_i\xi_i^* + \frac{d\xi_i}{dt}$. The solution to this differential equation is

$$c_i'(t) = \frac{\mu_0 M_s h_{mw}}{\upsilon_2 - \upsilon_1}\left[e^{\upsilon_1 t}\left(\frac{D}{j(\omega_{mw}-\omega_i)-\upsilon_1} - \frac{\gamma}{-j(\omega_{mw}-\omega_i)-\upsilon_1}\right) + e^{\upsilon_2 t}\left(\frac{-D}{j(\omega_{mw}-\omega_i)-\upsilon_2} + \frac{\gamma}{-j(\omega_{mw}-\omega_i)-\upsilon_2}\right)\right] \tag{15}$$

where $\upsilon_1 = \frac{1}{2}\left(2\Gamma_i + \sqrt{4\Gamma_i^2 - 4\lambda_i}\right)$, $\upsilon_2 = \frac{1}{2}\left(2\Gamma_i - \sqrt{4\Gamma_i^2 - 4\lambda_i}\right)$, and $D = \gamma(i\Omega_i - \Gamma_i + iR_i)$.

The total oscillation amplitude of each of the dots at each moment is the summation of all the oscillation modes containing that dot, $\left[c_T'(t)\right]_{N\times N} = \sum_{i=1}^{N_m} c_i'(t)[K_i]_{N\times N}$. The reversal for the dot at the row $m$ and the column $n$, at the time moment of $\tau$ can be deterministic if $\left| Re\left[ c'_{T,mn}(\tau) \right] \right| \geq 1$. Figure 6(b) demonstrates the analytical results for the possible frequencies for the reversal of only the center dot for all the initial states and different values of $H_{dc}$, in the cases that the reversal happens less than 30 ns, where $h_{mw} = 400$ Oe and $N = 3$. Figure 6(b) can be thought as the counterpart of the Fig. 4(a). The discrepancies between the analytical results and the numerical simulation results are due to the fact that for the sense of simplicity, we assume $H^x_{d,i}$, $H^y_{d,i}$ and $H^z_{d,i}$ to be constant during the dynamics, while in reality these components change with the growth of the amplitude of the oscillation modes.

## Discussion

As mentioned earlier, the Zeeman energy due to the localized dc and microwave fields in the excitation area forms an energy well that rules out the possibility of reversal in the outer dots. Due to zero $H_{dc}$ in the surrounding dots, their resonance frequencies of outer dots are very different from those of the dots inside the excited area, therefore, energy transmission from the inside to outside dots is negligible. However, the effect of surrounding dots outside the excitation area on the reversal process of the targeted dot cannot be ignored, since they act as sources of stray fields. A non-uniform stray field from the outer dots causes the change in the reversal spectrum. The stray field generated by the surrounding dots in the targeted dots matrix has a spatial pattern which depends on the magnetization status of the surrounding dots. We can write this stray field in the targeted matrix as $[H_s]_{N \times N} = \sum_{i=1}^{N_m} a_i [K_i]_{N \times N}$, where $N_m$ is the number of possible modes for a particular initial state of a N×N matrix. Therefore by extracting $a_i$ for a given $[H_s]_{N \times N}$, and knowing the effect of the value of $a_i$ on the reversal spectrum, $H_{dc}$, the excitation microwave frequency ($\omega_{mw}$), and the microwave amplitude ($h_{mw}$) can be adjusted to achieve the targeted reversal.

Another issue is the robustness against the initial magnetization direction and dispersion in the anisotropy field. As the microwave field is assumed to be applied in the x direction, the transfer of energy from microwaves to the dynamic modes depends on the azimuthal angle ($\varphi_{init}$) of the initial magnetization of the dots, and it can alter the excitation frequencies and the time required for the reversal of the central dot, as shown in Fig. 7(a) with a fixed polar angle ($\theta_{init}$) of 2° from the +z direction. Despite wide variations of $\varphi_{init}$, the change of switching frequencies and the required excitation time is moderate. Figure 7(b) shows the effect of ±11° variations around $\varphi_{init}$ = 45° for two different cases of $\theta_{init}$ = 4.5° and 5.5°. The excitation frequency and time required for the reversal of the central dot is relatively robust for slight variation of the initial magnetization direction around a predefined value ($\varphi_{init}$ = 45°). In practice, there could be a mechanism to set such an initial condition by applying a local or global dc field or by tilting the anisotropy direction during the material growth or annealing, preceding the microwave excitation.

Thermal agitation can perturb the anisotropy field. Figure 7(c) shows the effect of different levels of perturbation in the crystalline anisotropy constant, $K_u$ = $10^6$ J/m$^3$. The effect is negligible at several frequencies for $\Delta K_u$ up to the order of $10^4$ J/m$^3$. We have also studied a spatial distribution of the anisotropy field considering both a symmetric and an asymmetric case in the matrix. Figure 7(d) shows that even up to 10% of dispersion in $K_u$, there are several common frequencies to be utilized for switching. Thermal noise can be described as a Gaussian random variable in both time and space with a strength of $\sigma^2 = 2\alpha kT / \gamma M_s V$, where $k$ is the Boltzmann constant, $T$ is the temperature, and $V$ is the volume of the magnetic element[41]. At $T$ =

300 K and for the assumed material parameters, $\sigma/\mu_0$ = 10.9 Oe, which is negligible in comparison to the values of $h_{mw}$, $H_{dc}$ and $H_A$. The robustness shown in Fig. 7(b-d) indicates that the reversal is initiated through stable precessional modes, therefore the above thermal noise is negligible.

The proposed method of selective reversal needs to be carried out by sensing the initial magnetization state of all dots in the N×N matrix. The sensing can be done collectively. In addition, the total stray field from the outer dots needs to be measured by additional sensors, for example, using a 3×3 sensor matrix. Based on the initial pattern of the inner dots and the stray field from the outer dots, the proper microwave excitation (frequency, duration, and amplitude) as well as a dc field should be applied. The characteristics of proper microwave excitation should be preset for all the possible independent initial states and the stray field patterns from the outer dots, as what is done for a 3×3 isolated matrix in Fig. 4(a). After a reversal of the central dot in the N×N matrix, the write/read head should move as such the excitation area is shifted by $w + w_d$ (Fig. 1(a)). Afterwards, the reversal of the central dot of the new N×N matrix can be carried out. Alternatively, microwave striplines can be utilized for introducing defects in a magnonic crystal consisting of perpendicularly magnetized dot array. Each of the striplines can cover a matrix of dots. As the ground state of the array is known, the reversal of the central dot of each matrix can be carried out by applying a microwave current with a suitable frequency, duration, and amplitude to the striplines with a sufficient dc magnetic field.

The proposed method of the selective reversal of the center dot in a two dimensional N×N matrix can be expanded to the multilevel[42,43] matrices (N×N×P) for three dimensional recording. If each of the layers of the dots magnetostatically interacts with the other layers, an additional dimension of collective modes is added. As we are interested in the selective reversal of the center dot of each of the layers, introducing asymmetry in the excitation or magnetic characteristics among the layers is required. Figure 7(e) shows the frequencies for selective reversal of the center dot in either top (red color) or bottom (green color) layer of a 3×3×2 matrix, where the initial state of the dots is up. The collective sensing for such three dimensional matrices can be realized using the fact that, for each of the possible states of a 3D matrix, the combination of the stray field in the x and y directions for an asymmetric point relative to the matrix is non-degenerate (Fig. 7(f,g)). The measured stray field can be compared with an existing table to relate the sensed values to the possible states of the matrix. It is also possible to reconstruct the magnetization states without the comparing table. The magnetic dot $j$ generates the stray field of $\vec{H}_{j,i} = \frac{1}{4\pi}\left[\frac{3\vec{R}_{j,i}(\vec{M}_j \cdot \vec{R}_{j,i})}{|\vec{R}_{j,i}|^5} - \frac{\vec{M}_j}{|\vec{R}_{j,i}|^3}\right]$, at the sensing point of $S_i$ (Fig. 7(h)).

Assuming $\vec{M}_j = M_s m_{z,j} \hat{z}$, $\vec{H}_{j,i}$ in the x direction is $H_{x,j,i} = m_{z,j} A_{j,i}$, where $A_{j,i} = \frac{M_s}{4\pi}\left[\frac{3R_{x,j,i} R_{z,j,i}}{|\vec{R}_{j,i}|^5}\right]$. Therefore, the total stray field generated from all the dots of the matrix in the x direction at $S_i$ is $H_{T,x,i} = \sum_{j=1}^{N \times N \times N'} A_{j,i} m_{z,j}$. Consequently, by sensing N×N×P non-degenerate points in the space, and knowing the position of the dots and the sensing points, we can calculate the magnetization of the dots using

$$\begin{bmatrix} m_{z,1} \\ \vdots \\ m_{z,N^2 \times N'} \end{bmatrix} = \begin{bmatrix} A_{1,1} & \cdots & A_{N^2 \times N',1} \\ \vdots & \ddots & \vdots \\ A_{1,N^2 \times N'} & \cdots & A_{N^2 \times N',N^2 \times N'} \end{bmatrix} \begin{bmatrix} H_{T,x,1} \\ \vdots \\ H_{T,x,N^2 \times N'} \end{bmatrix}. \tag{16}$$

In summary, we propose a method for the magnetization reversal of a single nanodot by excitation in a matrix of dots. The method utilizes the microwave assisted magnetization reversal considering the inherent collective modes of the magneto-statically interacting dots. We demonstrate numerically the possibility of reversal of only one dot in a matrix of dots from any independent initial magnetization state with out-of-plane crystalline anisotropy. In addition, we discuss in more details the theory behind the proposed selective reversal. The proposed method can be utilized in microwave assisted switching for storage densities as high as 10 Tb/in$^2$, with write heads suitable for densities of 1.2 Tb/in$^2$, as well as in defect mode magnonic crystals in order to tune their band-structures.

## Methods

**Micromagnetic simulations.** We performed the numerical simulations using OOMMF. The simulated structures were matrices of cubic dots of 10 nm×10 nm×10 nm with inter-spacing of 10 nm in all the three directions (inter-distance in the z direction is only relevant to the 3D matrix, Fig. 7(f)). The structures were divided in 2.5 nm × 2.5 nm × 2.5 nm cells. For all the simulated structures, we used $M_s = 10^6$ A/m, the uniaxial crystalline anisotropy in the $z$ direction with the anisotropy constant of $K_u = 10^6$ J/m$^3$, the exchange constant of A = $10^{-11}$ J/m, and the Gilbert damping constant $\alpha = 0.005$. The applied Zeeman field consists of dc part ($H_{dc}$) in the −z direction, and a microwave field ($h_{mw}$) in the $x$ direction.

**Acknowledgements**: This research is supported by the National Research Foundation, Prime Minister's Office, Singapore under its Competitive Research Programme (CRP Award No. NRF-CRP 4-2008-06).

**Author contributions:** M.E, C.B., and H.Y. planned the study. M.E. carried out simulations and modeling. M.E. and H.Y. wrote the manuscript. H.Y. supervised the project.

**Additional information**

**Competing financial interests**: The authors declare no competing financial interests.

**Figure Captions**

**Fig. 1. Reversal patterns in a 3×3 matrix induced by excitation of the collective modes.** (a) The schematic of a dot matrix, the configuration of fields, and the excitation area. The dot number is defined for a 3×3 matrix. (b) Some of the realized reversal pattern for a 3×3 matrix of dots with respect to the excitation time and frequency, where $H_{dc}$ = 14.5 kOe and $h_{mw}$ = 400 Oe. (c) Each color stands for a magnetization pattern defined in the bottom of the graph.

**Fig. 2. Relation between the reversal patterns and the oscillation amplitudes of the dots.** (a) Excitation time and frequency dependence for some of the realized reversal patterns, with two opposite corners initially magnetized to the −z direction (marked in red). The magnetic configuration of each dot is defined in the graph. (b) $M_z^2$ of all the nine dots versus frequency. Frequencies in which the dot number 5 has low values of $M_z^2$ are marked with the black arrows. (c) The reversal sequence for $f$ = 10.75 GHz. (d) $M_z^2$ for all the dots at $f$ = 10.75 GHz with respect to time. Colored circles stand for the patterns shown in (a) and (c). (e) The normalized oscillation power ($P_o$) at different frequencies and dots for $h_{mw}$ = 50 Oe. (f) The oscillation phase of the dots with respect to that of dot 5 at $t$ = 3 ns for $h_{mw}$ = 50 Oe.

**Fig. 3. Relation between the applied fields and the reversal characteristics.** (a) The number of possible reversal patterns for different values of $H_{dc}$. (b) Frequencies for achieving the reversal of only the central dot (dot 5) for various $H_{dc}$. (c) The number of possible reversal patterns for different values of $h_{mw}$. (d) Frequencies for achieving the reversal of only the central dot for various $h_{mw}$.

**Fig. 4. Reversal from independent initial states.** (a) The frequencies that the reversal of the central dot (dot 5) is possible for the 51 possible independent initial states. (b) The $N_s/N_p$ for the 51 possible independent initial states. Some initial states are depicted beside their corresponding points. The applied fields are $H_{dc}$ = 14 kOe and $h_{mw}$ = 400 Oe.

**Fig. 5. Reversal spectrum for a 7×7 matrix.** (a) Excitation time and frequency dependence for some of the realized reversal patterns for a 7×7 matrix of dots. The initial states is all the dots being magnetized to the z direction. (b) The states corresponding to each color are depicted. The applied fields are $H_{dc}$ = 14.5 kOe and $h_{mw}$ = 450 Oe.

**Fig. 6. Results of the analytical model for a 3×3 matrix.** (a) The frequencies in which the stationary oscillation amplitude of the center dot (dot 5) is larger than the other dots for the 51 possible independent initial states. The graph includes results for $H_{dc}$ from 14 to 19 kOe with a step of 0.5 kOe. (b) The frequencies in which only the center dot (dot 5) reverses in less than 30 ns for the 51 possible independent initial states. The graph includes results for $H_{dc}$ from 14 to 16 kOe with a step of 0.5 kOe. In (a) and (b), the colors other than background dark blue, refer to $H_{dc}$ values.

**Fig. 7. Considerations for applications.** (a) The frequencies that the reversal of the central dot (dot 5) is possible for different azimuthal angles ($\varphi_{init}$) of the initial magnetization, while the polar angle ($\theta_{init}$) was fixed at 2°. $t_{min}$ is the minimum excitation time required to achieve the reversal of the central dot. The darkest blue is the background. (b) Frequencies that the reversal of the central dot is possible for $\varphi_{init}$ = 45, 45±11° at $\theta_{init}$ = 4.5 or 5.5°. (c) The frequencies that the reversal of the central dot is possible for different levels of perturbation in the anisotropy constant with $\varphi_{init}$ = 45° and $\theta_{init}$ = 5°. (d) The frequencies that the reversal of the central dot is possible for different levels of dispersion in the anisotropy constant, for both a symmetric (S) and an asymmetric (AS) case with $\varphi_{init}$ = 45° and $\theta_{init}$ = 5°. In the case of S, the anisotropy value of dots 1, 3, 7, and 9 is modified as $K_u+\Delta K_u$. For the AS cases, the anisotropy values of dots 1, 2, 5, and 7 are modified as $K_u+\Delta K_u$, $K_u+0.5\Delta K_u$, $K_u-0.5\Delta K_u$, and $K_u-\Delta K_u$, respectively. (e) The excitation frequencies and times that the reversal of the central dot of the bottom (top) layer is possible for the initial state of all the dots being up are shown with green (red). $K_{u,1} = 0.6\times10^6$ J/m$^3$ and $h_{mw,1} = 270$ Oe ($K_{u,2} = 1\times10^6$ J/m$^3$ and $h_{mw,2} = 300$ Oe) are the anisotropy and microwave field for the bottom (top) layer of a 3×3×2 matrix of dots, respectively. (f) Schematic of the relative position of a sensor in a 3D matrix of dots. (g) An asymmetric sensing point from the top view (green square is the sensor position). (h) A relative position of the dot $j$ and a sensing point $S_i$.

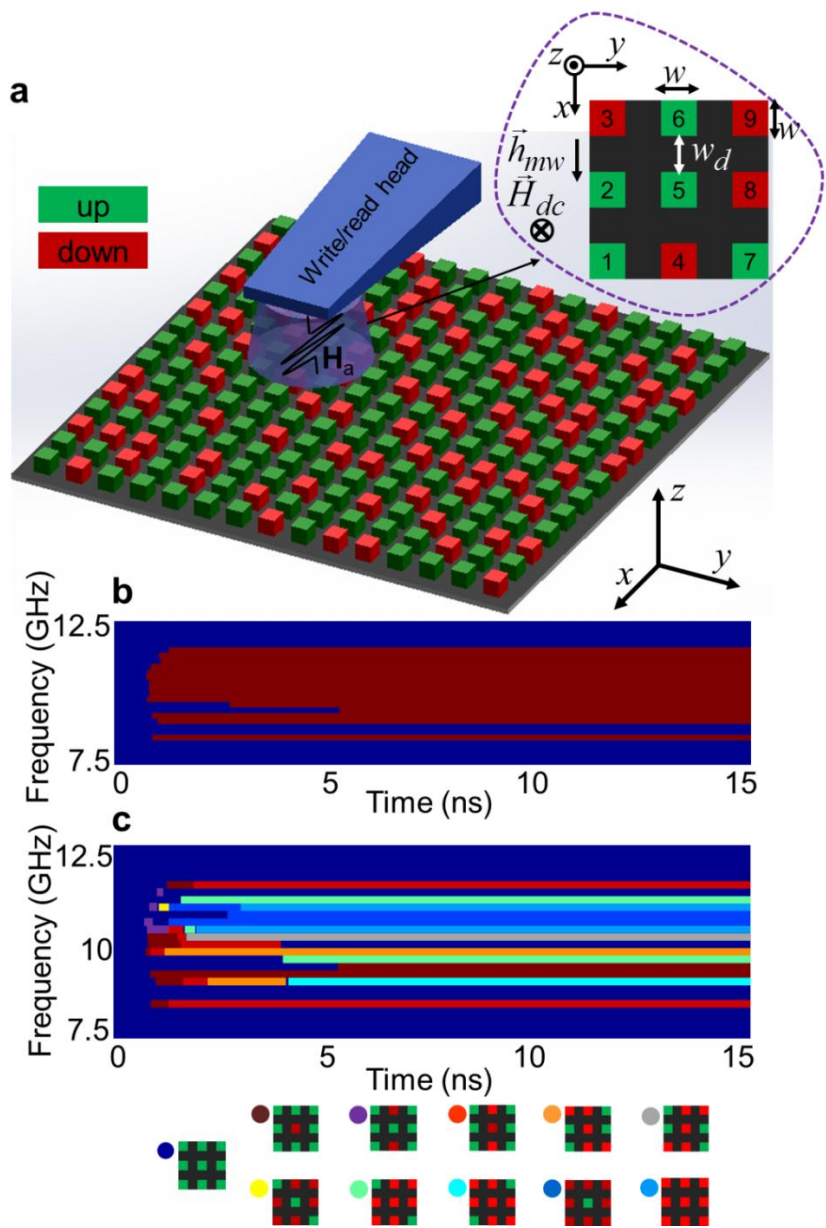

**Fig. 1**

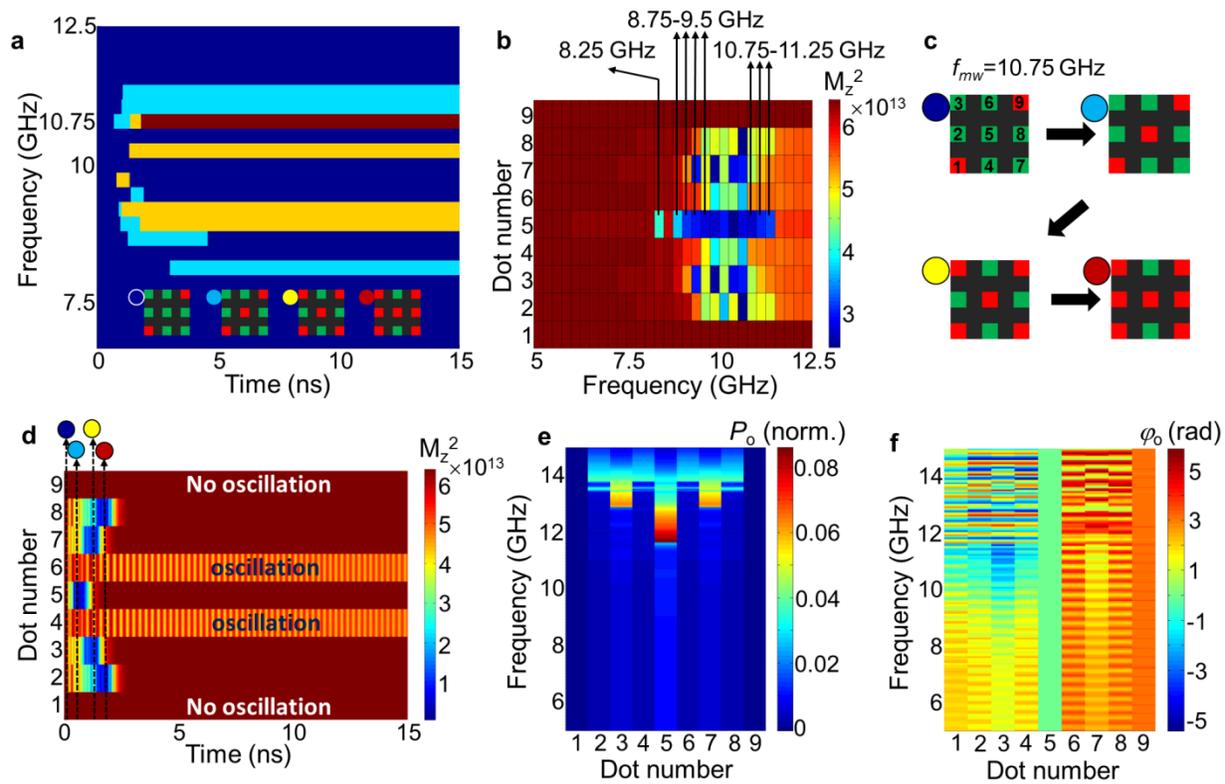

**Fig. 2**

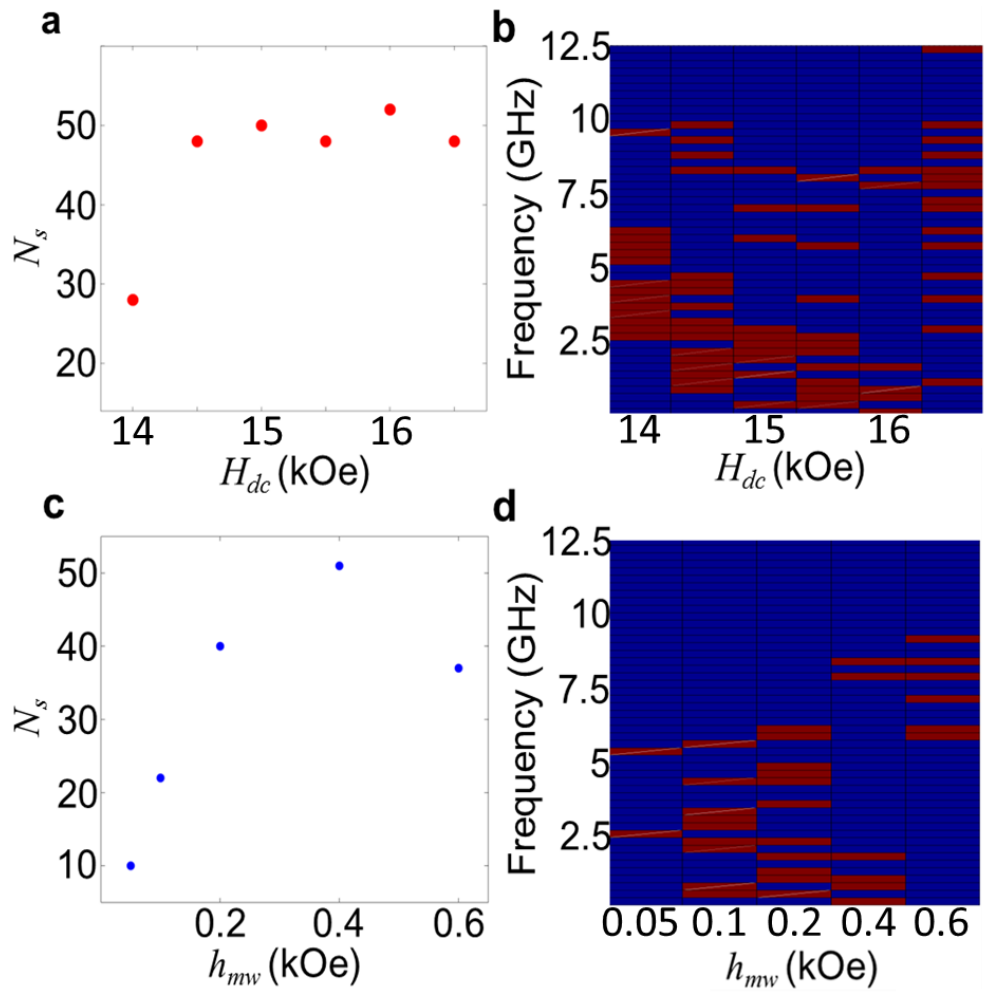

**Fig. 3**

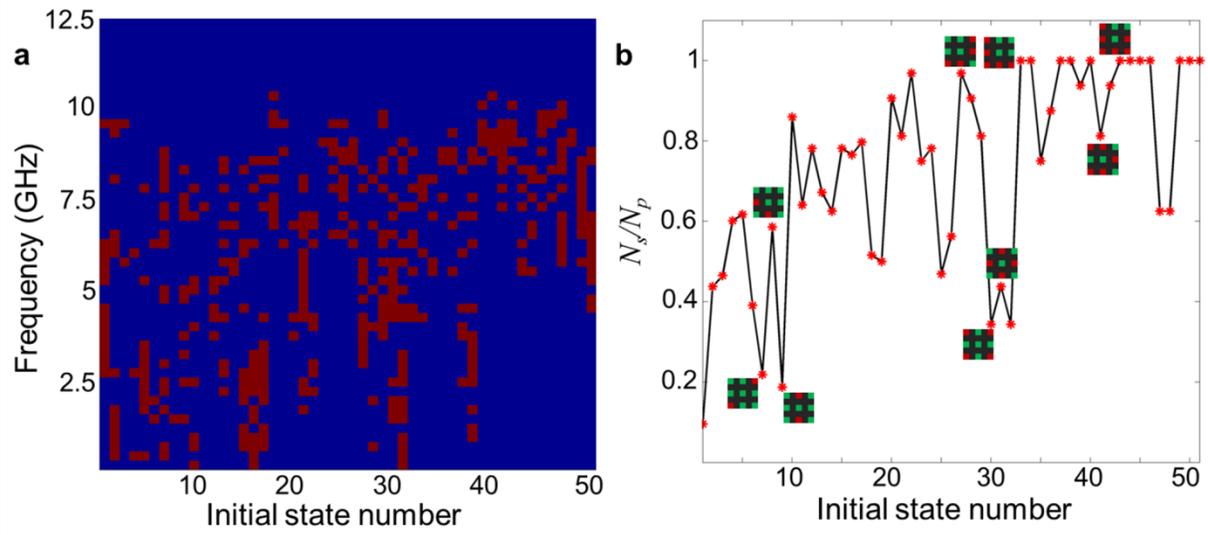

**Fig. 4**

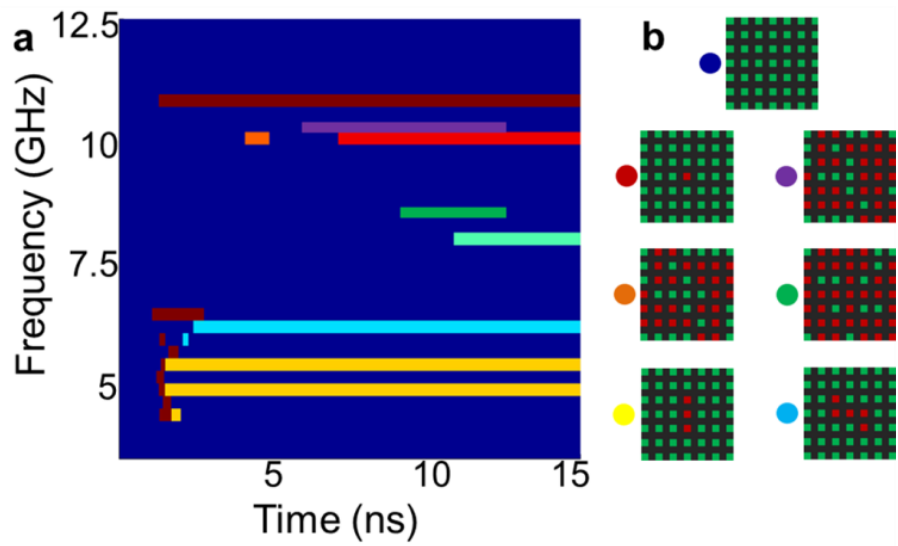

**Fig. 5**

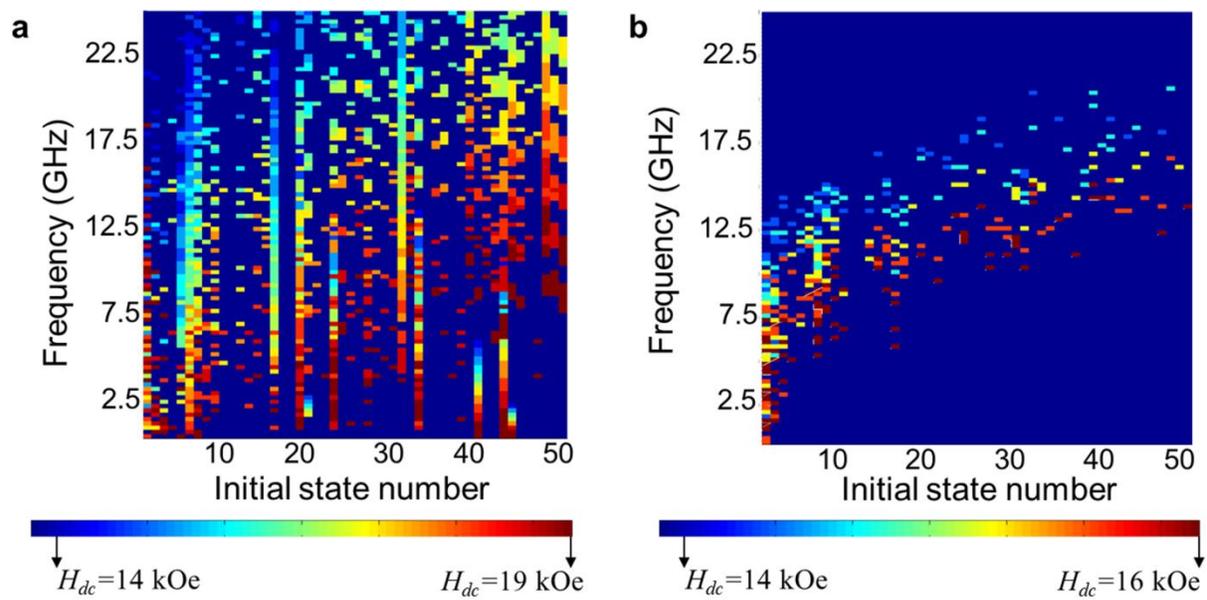

**Fig. 6**

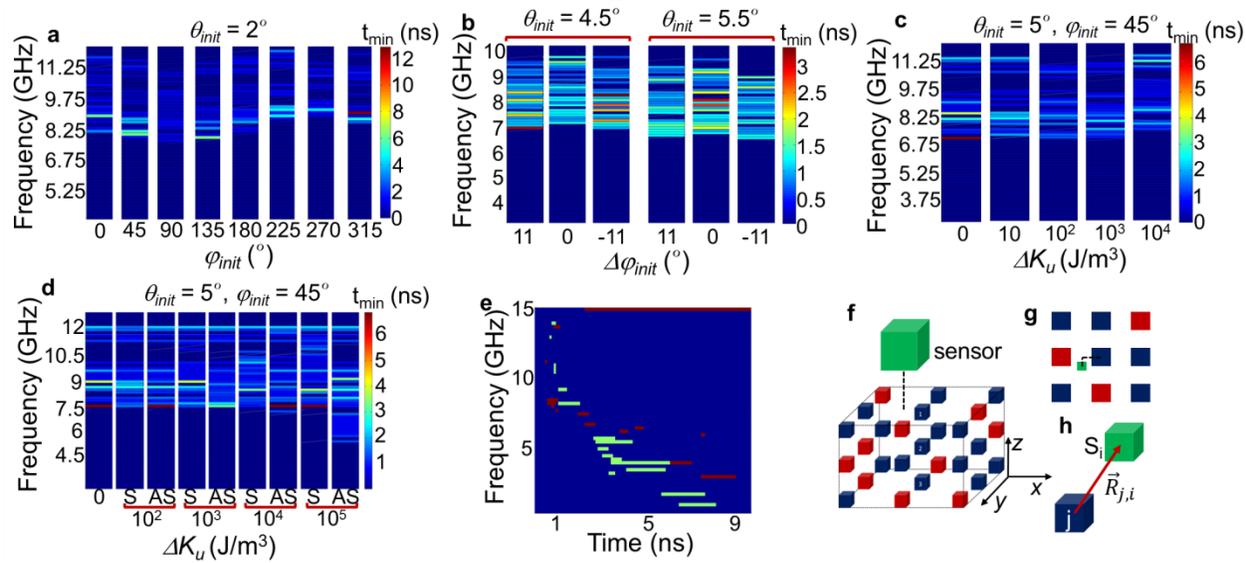

**Fig. 7**